\documentclass[preprint,11pt,3p]{elsarticle}
\usepackage{graphicx}
\usepackage{dcolumn}
\usepackage{bm}
\usepackage{hyperref}
\usepackage{amssymb}
\usepackage{hyperref}
\RequirePackage[normalem]{ulem} 
\RequirePackage{color}\definecolor{RED}{rgb}{1,0,0}\definecolor{BLUE}{rgb}{0,0,1} 

\begin{document}
\newcommand{\be}{\begin{equation}}
\newcommand{\ee}{\end{equation}}
\newcommand{\ba}{\begin{eqnarray}}
\newcommand{\ea}{\end{eqnarray}}
\newcommand{\Gam}{\Gamma[\varphi]}
\newcommand{\Gamm}{\Gamma[\varphi,\Theta]}
\thispagestyle{empty}

\title
{
 Two-electron entanglement in a two-dimensional isotropic harmonic trap: Radial
correlation effects in  the low density limit. }
\author{ Przemys\l aw Ko\'scik,
Institute of Physics,  Jan Kochanowski University\\
ul. \'Swi\c{e}tokrzyska 15, 25-406 Kielce, Poland}
\ead{koscik@pu.kielce.pl}

 \begin{abstract} We study  a two-dimensional system of two Coulombically interacting
electrons   in an external  harmonic confining potential. More
precisely, we present calculations for the  singlet ground-state of
the system. We explain the nature of the degeneracy   in the
spectrum of the reduced density matrix and provide detailed results
for the dependencies of the entanglement on the interaction
strength. Among other features, it is indicated that  in the limit
of an infinitely strong  interaction, the
  angular  and radial correlations are asymptotically  independent of  each other.
 Appearing in this limit, the pure
radial   correlation effects are quantitatively investigated for the
first time.

 \end{abstract}


\maketitle

\section{Introduction}

Studies of quantum systems trapped in confining potentials have
attracted attention because of their possible applications in
quantum information technology~\cite{QCdots}. The simplest candidate
for studying entanglement properties is the  system composed of two
Coulombically interacting electrons  in an external harmonic
potential,  which serves well as a model of a quantum dot
(QD)\cite{fab}. In this Letter we consider   a two-dimensional (2D)
system with a  Hamiltonian
\begin{eqnarray} \hat{H}=\sum_{i=1}^2[{\textbf{p}_{i}^2\over 2 m_{*}}
+ {\omega^2m_{*} \over 2} r_{i}^2]
 +{e^2\over
\varepsilon_{*}|\textbf{r}_{2}-\textbf{r}_{1}|},~~~~\label{ham}\end{eqnarray}
where $m_{*}$ is the effective electron mass and $\varepsilon_{*}$
is the effective dielectric constant. This system  was considered in
various theoretical contexts
 \cite{tautB,Yan,MatulisPeeters,Puente,has}
but  to the best of our knowledge its entanglement properties  have
not been studied extensively in the literature.

Since the Hamiltonian (\ref{ham}) is spin independent its stationary
states possess the form \be
\Psi(\zeta_{1},\zeta_{2})=\psi(\textbf{r}_{1},\textbf{r}_{2})\varrho(s_{1},s_{2}),\label{fum}\ee
where $\varrho$ denotes a spin function. After the  scaling
$\textbf{r}\mapsto \sqrt{2\hbar\over {{m^{*}\omega}}}\textbf{r}$,
$E\mapsto {\hbar \omega E\over {2}}$,
 the  Schr\"{o}dinger equation  takes the form
\be
H\psi(\textbf{r}_{1},\textbf{r}_{2})=E\psi(\textbf{r}_{1},\textbf{r}_{2}),\label{EOM}
\ee with
\begin{eqnarray} H= \sum_{i=1}^2[-{1\over
2}\triangle_{\textbf{r}_{i}}+2 r_{i}^2]
 +{g\over
|\textbf{r}_{2}-\textbf{r}_{1}|} \label{hamrt},\end{eqnarray} where
the dimensionless coupling
$g={e^2\over\varepsilon^{*}}\sqrt{{2m^{*}\over \omega\hbar^{3}}}$
represents the ratio of Coulomb interaction strength, which  is
related to the so-called Wigner parameter $R_{w}$ of Ref.\ \cite{Yan}   by
$g=\sqrt{2} R_{w}$.

Introducing the centre-of-mass (cm.) $\textbf{R}={1\over
2}(\textbf{r}_{1}+\textbf{r}_{2})$  and relative (rel.)
$\textbf{r}=\textbf{r}_{2}-\textbf{r}_{1}$ coordinates, the
Hamiltonian (\ref{hamrt}) is separated into
 $H=H^{\textbf{R}}+H^{\textbf{r}}$, where the cm.\ Hamiltonian $H^{\textbf{R}}={-\nabla_{R}^2/
4}+4R^2\label{cm}$ is exactly solvable. The problem is thus reduced
to the  Schr\"{o}dinger equation for the rel.\ motion
$$H^{\textbf{r}}\psi^{r}=\varepsilon^{\textbf{r}}\psi^{r},$$
described by the Hamiltonian $H^{\textbf{r}}=-\nabla_{r}^2 +
r^2+{g\over r}$. For the sake of simplicity we restrict our
investigation to the study of the singlet ground-state. The total
wavefunction is given by
     \begin{eqnarray}
\Psi(\zeta_{1},\zeta_{2})&=&\psi(r,R)\varrho_{S}(s_{1},s_{2})=\nonumber\\
&=&{ 2\over \sqrt{\pi}}e^{-2R^2}{1\over \sqrt{2\pi}}{u^{r}(r)\over
\sqrt{ r} } \varrho_{S}(s_{1},s_{2}), \label{wfd}\end{eqnarray}
where the unknown function $u^{r}$
($\int_{0}^{\infty}[u^{r}]^2dr=1$) is the lowest-energy
eigenfunction of the rel.\ radial equation \be [-{d^2\over
dr^2}-{1\over 4r^2}+ r^2+{g\over
r}]u^{r}(r)=\varepsilon^{\textbf{r}}u^{r}(r),\label{radial}\ee and
$\varrho_{S}={\frac{1}{\sqrt{2}}}(\alpha(1)\beta(2)-\alpha(2)\beta(1))$
is the spin singlet function,  where $\alpha(\beta)$ denotes the up
(down), $\uparrow(\downarrow)$, spin. We note that
$r=\sqrt{r_{1}^2+r_{2}^2-2r_{1}r_{2}cos(\varphi_{2}-\varphi_{1})}$
and
$R=\sqrt{r_{1}^2+r_{2}^2+2r_{1}r_{2}cos(\varphi_{2}-\varphi_{1})}/2$,
so that in a correlated description $\psi$ depends explicitly on
$r_{1}$, $r_{2}$ and   the inter-electronic angle coordinate
 $\theta=\varphi_{2}-\varphi_{1}$.

In this Letter we will analyse   the entanglement contained in the
state $|\Psi\rangle$ (\ref{wfd}). Special attention is given to
performing both the partial-wave expansion and the Schmidt
decomposition of $\psi$. We make a detailed investigation of the
nature of the degeneracy
 in the spectrum of the reduced density matrix. We discuss  the effect of interaction  on various
characteristics such as the collective occupancies, the
participation ratio,
 and
the Slater rank as well, over a wide range of values of  $g$.
 In particular, we   show  that
only the Schmidt modes with nodeless radial  functions contribute
significantly  to the total wavefunction. Importantly, we give an
indication that in the $g\rightarrow\infty$ ($\omega\rightarrow 0$)
limit the global
 correlation is  fully separable into purely angular and radial correlations.
Moreover, in the regime of  $g\rightarrow\infty$ we investigate for the first time  the
 role played by the  radial correlation
 on  the electron pair.

 The Letter  is arranged as follows. In Section \ref{sec2} we
discuss the
 characteristics of the system under consideration.
 Section \ref{res} is devoted to the results and  a
brief summary of our conclusions is provided in Section \ref{summ}.

\section{Entanglement characteristics}\label{sec2}
 A  tool to investigate two-body correlations
 is the reduced density matrix (RDM)  defined as \cite{redu}
\be \rho_{red}(\zeta,\zeta^{'})= tr_{\zeta_{2}}(|\Psi\rangle\langle
\Psi|), \label{RDM}\ee where $tr$ denotes a trace taken over one of
the electrons, and $\Psi$ is the total two-electron wavefunction. As
$\Psi$ is given  by Eq.~(\ref{fum}), the RDM factors into spatial
and spin components \be \rho_{red}(\zeta,\zeta^{'})=
\rho(\textbf{r},\textbf{r}^{'})\rho_{S}(s,s^{'}), \ee where the
spatial RDM is given by
$\rho(\textbf{r},\textbf{r}^{'})=\int\psi^{*}(\textbf{r},\textbf{r}_{2})\psi(\textbf{r}^{'},\textbf{r}_{2})
d\textbf{r}_{2}$. In particular, for the singlet states, the spin
part of the RDM
  is a  matrix $\rho_{S}=diag(1/2 ,1/2)$. It is well known that for identical
fermions the bi-partite pure state $|\Psi\rangle$ can be expressed
as a combination of the Slater determinants made out of one-particle
spin-orbitals in which the RDM (\ref{RDM}) is diagonal \cite{Ghirardi}.
  The number of  non-zero  expansion  coefficients  in the Slater decomposition
  is called the Slater rank (SR)
   and a pure fermion state is entangled
if, and only if, its Slater rank is larger than $1$.
The   entanglement depends on the whole spectrum of  the RDM and
many ways of measuring its amount have been developed in the
literature \cite{par,lin,batle,vn1}. It is interesting to point out
that most of the popular entanglement measures  are  functions of
the purity of the RDM, $tr\rho^2_{red}$. The best known are the
participation ratio \cite{par} and the linear entropy \cite{lin}.
Here we consider 
the  participation ratio,
 \be \textbf{R}=[tr \rho_{red}^2]^{-1},\label{part}\ee
  which  approximately counts the   number of Slater orbitals
 actively involved in the Slater expansion of $\Psi$.
The effective Slater rank $SR$ is defined by $SR=\textbf{R}/2$,
since the Slater determinant is made up of  two different spin
orbitals.   The larger the value of
 $SR$, the higher the entanglement or quantum correlation. It is worth stressing at this
 point that
 the linear correlation entropy $\textbf{L}$, which is the
  popular measure of entanglement in pure states \cite{lin0,lin1,lin20,lin3,lin4},  is
related to  $\textbf{R}$ via $\textbf{ L}=1-1/\textbf{R}$.


\subsection{The Slater-Schmidt decomposition}
The Slater decomposition of the total wavefunction of (\ref{fum})
 can easily be  inferred from the Schmidt decomposition of  the spatial part $\psi$.
In a previous work \cite{kos}, we have shown that
  for any 2D system
 the real spatial
wavefunction that depends only on  $r$ and $R$,
$\psi(r,R)\equiv\psi(r_{1},r_{2},cos\theta)$,  can be expanded as
\begin{eqnarray}
\psi(r,R)=\sum_{m=-\infty...\infty}{A_{m}(r_{1},r_{2})\over
\sqrt{r_{1}r_{2}}}{e^{im\varphi_{1}}\over
\sqrt{2\pi}}{e^{-im\varphi_{2}}\over
\sqrt{2\pi}}=\nonumber\\=\sum_{m=-\infty...\infty\atop
s=0,\infty}\kappa_{s,m}u_{s,m}(\textbf{r}_{1})u_{s,m}^{*}(\textbf{r}_{2}),
\label{ddd}\end{eqnarray} where $A_{m}=A_{-m}$,
\begin{eqnarray}
 A_{m}(r_{1},r_{2})={\sqrt{r_{1}r_{2}}\over 2\pi}\int_{0}^{2\pi}\int_{0}^{2\pi}\psi(r,R){e^{-im\varphi_{1}}}{e^{im\varphi_{2}}}d\varphi_{1}d\varphi_{2}=\nonumber\\=\sqrt{r_{1}r_{2}}\int_{0}^{2\pi}\psi(r_{1},r_{2},cos\theta)cos(m\theta) d\theta
\label{rrr},\end{eqnarray}
 and  \be
A_{m}(r_{1},r_{2})=\sum_{s=0}^{\infty}\kappa_{s,m}\chi_{s,m}(r_{1})\chi_{s,m}(r_{2})\label{schaaa},\ee
\be u_{s,m}(\textbf{r})={\chi_{s,m}(r)\over
\sqrt{r}}{e^{im\varphi}\over \sqrt{2\pi}}\label{pp}.\ee The radial
orbitals $\chi_{s,m}(r)$  and  the coefficients $\kappa_{s,m}$ are
determined by the integral equation \be \int_{0}^{\infty}
A_{m}(r_{1},r_{2})\chi_{s,m}(r_{2})dr_{2}=\kappa_{s,m}\chi_{s,m}(r_{1}),
\label{aa}\ee and the family $\{u_{s,m}(\textbf{r})\}$ forms a
complete and orthogonal set ($\langle u_{sm}| u_{s^{'}m^{'}}\rangle=
 \int_{0}^{\infty}\int_{0}^{2\pi}r u_{sm}^{*} u_{s^{'}m^{'}}dr
d\varphi$
$=\delta_{mm^{'}}\int_{0}^{\infty}\chi_{sm}\chi_{s^{'}m}dr$$=
\delta_{mm^{'}}\delta_{ss^{'}}$), which means that the latter
expansion appearing in Eq.~(\ref{ddd}) represents  the Schmidt
decomposition of $\psi$.
Since the components $A_{m}$ and $A_{-m}$   are the same,
Eq.~(\ref{ddd}) can be written  in a partial form  \be
\psi(r,R)=\nonumber\\{A_{0}(r_{1},r_{2})\over 2\pi\sqrt{ r_{1}r_{2}}
}+\sum_{l=1}^{\infty}{A_{l}(r_{1},r_{2})\over
\sqrt{r_{1}r_{2}}}{cos(l \theta)\over \pi}\label{partwave}, \ee and
$\chi_{s,m}=\chi_{s,-m}$, $\kappa_{s,m}=\kappa_{s,-m}$. The latter
means that the Schmidt decomposition  possesses a degenerate
spectrum, and so
 it fails to be unique \cite{Ghirardi}.
 For the sake of completeness,  we
give below another from Eq.~(\ref{ddd})
  form of the Schmidt decomposition  of $\psi$.
 To begin with, we extend  the results of Ref.\  \cite{Ghirardi}  to the case with
  more than one point of double degeneracy.
  Accordingly, from the orbitals $u_{s,l}$ and $u_{s,-l}$ (\ref{pp}) ($l>0$), that
  correspond
  to the same Schmidt coefficient $\kappa_{s,l}$, we define the new orbitals as
$$\nu_{s,l}(\textbf{r})={u_{s,l}(\textbf{r})+u_{s,-l}(\textbf{r})\over
\sqrt{2}},
\upsilon_{s,l}(\textbf{r})=i{u_{s,-l}(\textbf{r})-u_{s,l}(\textbf{r})\over
\sqrt{2}},$$ and for $l=0$ as $\nu_{s,0}={\chi_{s,0}(r)\over
\sqrt{2\pi}\sqrt{r} }=u_{s,0}$, $\upsilon_{s,0}=0$. The angular
parts of the   new Schmidt modes  can be obtained analytically.
After some tedious algebra, one arrives at
\begin{eqnarray} \nu_{s,l}(\textbf{r})={\chi_{s,l}(r)\over \sqrt{r}}{cos(l\varphi)\over \sqrt{\pi}},
\upsilon_{s,l}(\textbf{r})={\chi_{s,l}(r)\over
\sqrt{r}}{sin(l\varphi)\over \sqrt{\pi}}.\end{eqnarray} The family
$\{\nu_{s,l},\upsilon_{s,l}\}$ forms a complete and
 orthonormal set, since  one can easily check that
 $\langle\nu_{sl}|\nu_{s^{'}l^{'}} \rangle=\delta_{ss^{'}}\delta_{ll^{'}}$,
 $\langle\upsilon_{s,l}|\upsilon_{s^{'}l^{'}} \rangle=\delta_{ss^{'}}\delta_{ll^{'}}$,
 $\langle\nu_{s,l}|\upsilon_{s^{'}l^{'}} \rangle=0$.
   In terms of the  new orbitals,  the wavefunction takes the form
\begin{eqnarray}\psi(r,R)=\sum_{l,s=0}^{\infty}\kappa_{s,l} [\nu_{s,l}(\textbf{r}_{1}) \nu_{s,l}(\textbf{r}_{2})+\upsilon_{s,l}(\textbf{r}_{1})\upsilon_{s,l}(\textbf{r}_{2})]\label{sc1},\end{eqnarray}
which yields a Schmidt form different  from Eq.~(\ref{ddd}).
Consequently, there exist also two different forms of the Slater
expansion of (\ref{wfd}), which can easily be inferred from the
Schmidt decompositions (\ref{ddd}) and  (\ref{sc1}).

\subsection{The partial expansion of the RDM }\label{pr}

Now using Eq.~(\ref{ddd}) we  turn the spatial RDM into the partial
form
\begin{eqnarray}
\rho(\textbf{r},\textbf{r}^{'})={\rho_{0}(r,r^{'})\over 2 \pi\sqrt{
rr^{'}}} +\sum_{l=1}^{\infty}{\rho_{l}(r,r^{'})\over
\sqrt{rr^{'}}}{cos(l \theta^{'})\over \pi},\label{3e}\end{eqnarray}
wherein  $\theta^{'}=\varphi-\varphi^{'}$, and $\rho_{l}$ (the
$l$-matrix) is given by   \be \rho_{l}(r,r^{'})= \int_{0}^{\infty}
A_{l}(r,r_{2})A_{l}(r^{'},r_{2})dr_{2}\label{polko}.\ee
It can easily be  derived that the $l$-matrix has the expansion 
\be\rho_{l}(r,r^{'})=\sum_{s=0}^{\infty}\lambda_{s,l}\chi_{s,l}(r)\chi_{s,l}(r^{'})\label{expan},\ee
which represents nothing else but its Schmidt decomposition. The
eigenequation of the $l$-matrix  is thus of the form
 \be \int_{0}^{\infty} \rho_{l}(r,r^{'})\chi_{s,l}(r^{'})dr^{'}=\lambda_{s,l}\chi_{s,l}(r).\ee
Obviously, the spatial RDM (\ref{3e}) is diagonal  in the basis of
Schmidt modes (spatial natural orbitals) and its eigenvalues
$\lambda_{s,l}$ (occupancies) are related to $\kappa_{s,l}$  via the
formula $\lambda_{s,l}=\kappa_{s,l}^2$.
Moreover, all  occupancies but the ones with $l=0$ are doubly
degenerate so that the normalization condition gives
$\sum_{s=0}^{\infty}\lambda_{s,0}+2\sum_{l=1}^{\infty}\sum_{s=0}^{\infty}\lambda_{s,l}=1$.

The quantity of interest  is  the sum of the occupancies
corresponding to the $l$-matrix
$\eta_{l}=\sum_{s=0}^{\infty}\lambda_{s,l}$. Throughout this Letter
we will refer to  $\{\eta_{l}\}$ as the collective occupancies. They
can simply be calculated by $\eta_{l}= Tr\rho_{l}=\int_{0}^{\infty}
\rho_{l}(r,r)dr=||A_{l}||^2$,
 which is crucial since for their determination   the numerical
 diagonalizations of the matrices $\rho_{l}$ are not needed.
It is clear that, if  $\eta_{l}$ is vanishingly small compared with
the remaining collective occupancies,  we may expect that
 the $l$-partial wave  in Eq.~(\ref{partwave})  contributes  little.
Moreover,  $\eta_{0}$ might be seen as the fraction of electrons with
zero angular momentum, and $2\eta_{l}$ ($l>0$), as
the  fraction of  electrons with  fixed opposite angular momenta 
($\hbar l$ and   $-\hbar l$). This is particularly easy to
understand  when   rewriting the spatial wavefunction as
 a combination of
 symmetrized products of the modes $u_{s,l}$, $u_{s,-l}$ (\ref{pp}) \footnote{
$\psi(r,R)=\sum_{s=0}^{\infty}\kappa_{s,0}u_{s,0}(\textbf{r}_{1})u_{s,0}(\textbf{r}_{2})+\sum_{s=0,l=1}^{\infty}\sqrt{2}\kappa_{s,l}[{1\over\sqrt{2}}(u_{s,l}(\textbf{r}_{1})u_{s,-l}(\textbf{r}_{2})+u_{s,l}(\textbf{r}_{2})u_{s,-l}(\textbf{r}_{1}))]$}.

\subsection{The participation ratio }\label{pr}
The purity of  the RDM, $tr\rho^2_{red}$, of
 the singlet state
separates in $ {1\over 2}tr \rho^2$, where ${1\over 2}$ is related
to the spin degree of freedom. Moreover, in the case of  (\ref{3e}),
 $tr \rho^2$ can be decomposed as
\be tr \rho^2=tr \rho^2_{0}+2\sum_{l=1}^{\infty}tr
\rho^2_{l},\label{line22}\ee where \be tr
\rho^2_{l}=\int_{0}^{\infty}\rho^2_{l}(r,r)dr, \ee with \be
\rho^2_{l}(r_{1},r_{2})=\int_{0}^{\infty}
\rho_{l}(r_{1},r_{3})\rho_{l}(r_{3},r_{2}) dr_{3}.  \ee Accordingly,
in terms of (\ref{line22}) the participation ratio  of the singlet
state can be written as \be \textbf{R}=[{1\over 2}tr
\rho^2_{0}+\sum_{l=1}^{\infty}tr \rho^2_{l}]^{-1}.\label{line}\ee
From a practical point of view the above expansion appears to be a
convenient tool to determine the number $N_{l}$ of partial waves
that are needed to capture most of the electrons correlation. It
seems to be reasonable to define
  $N_{l}$ as the
minimal number of terms in the sum in   (\ref{line})   at which an
approximate value for   $\textbf{R}$ has its   integer
part equal to the  integer part of the exact value of $\textbf{R}$.

Clearly, the  effective  Schmidt number of non-zero coefficients in
the Schmidt decomposition  of ${A_{l}}$, Eq.~(\ref{schaaa}), is
determined by the   participation ratio of the normalized
$l$-matrix, \be \bar{\rho_{l}}(r,r^{'})={{\rho_{l}(r,r^{'})}\over
Tr\rho_{l}},\label{rdml7}\ee
 ($Tr \bar{\rho_{l}}=1$), i.e. $[tr \bar{\rho_{l}}^2]^{-1}$.

 \section{ Numerical results }\label{res}

\begin{figure}[h]
\begin{center}
\includegraphics[width=0.65\textwidth]{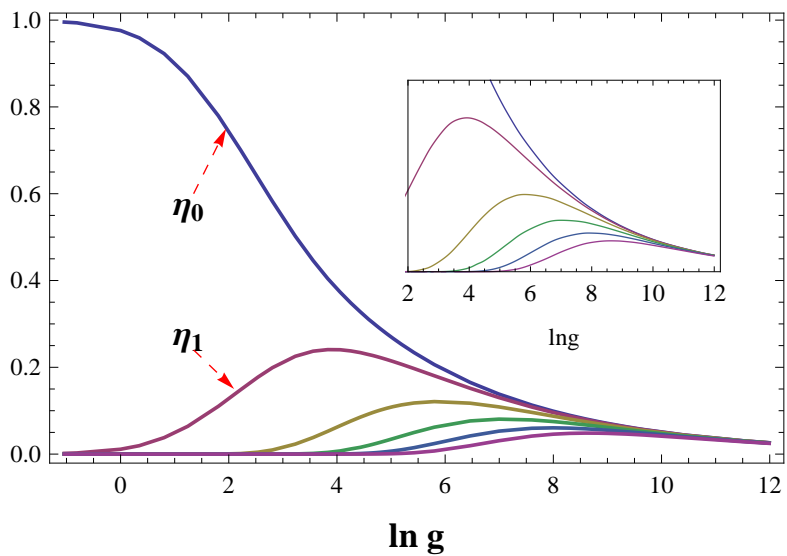}
\end{center}
 \caption{\label{fig25ee3f:beh}The dependence  of the six lowest collective occupancies
  $\eta_{l}$ $(l=0-5)$ of the lowest singlet state on  $\ln g $.
  The inset highlights their changes for large values of  $g$.
   }
 \end{figure}

\begin{figure}[h]
\begin{center}
\includegraphics[width=0.65\textwidth]{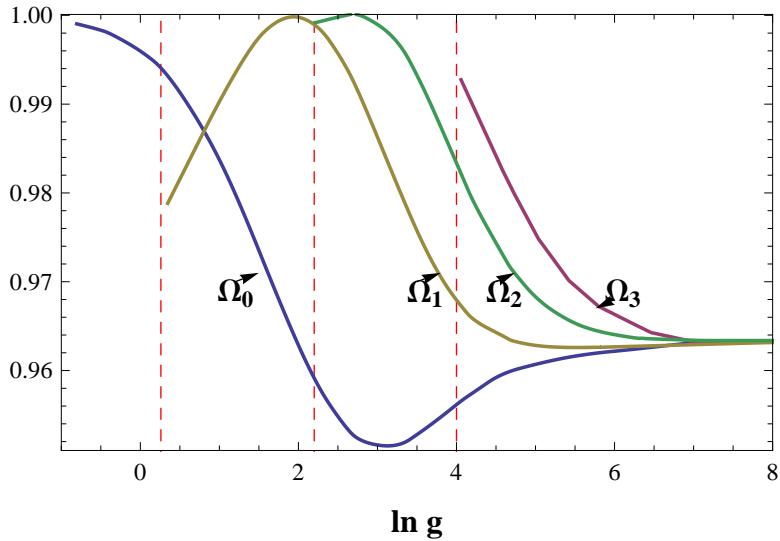}
\end{center}
 \caption{\label{fig2ffo65e3f1:beh}The  behaviour of    $\Omega_{l}$
   for $l=0-3$
    as a function of
   $\ln g$. The first vertical line (on the left) corresponds to the value of $g$  below which
the partial wave with $l=1$   hardly contributes to $\psi$   at all.
The second line concerns the same but for $l=2$,
and so on. The locations of the lines
 were   roughly  inferred from Fig.~\ref{fig25ee3f:beh}.
   }
 \end{figure}

\begin{figure}[h]
\begin{center}
\includegraphics[width=0.65\textwidth]{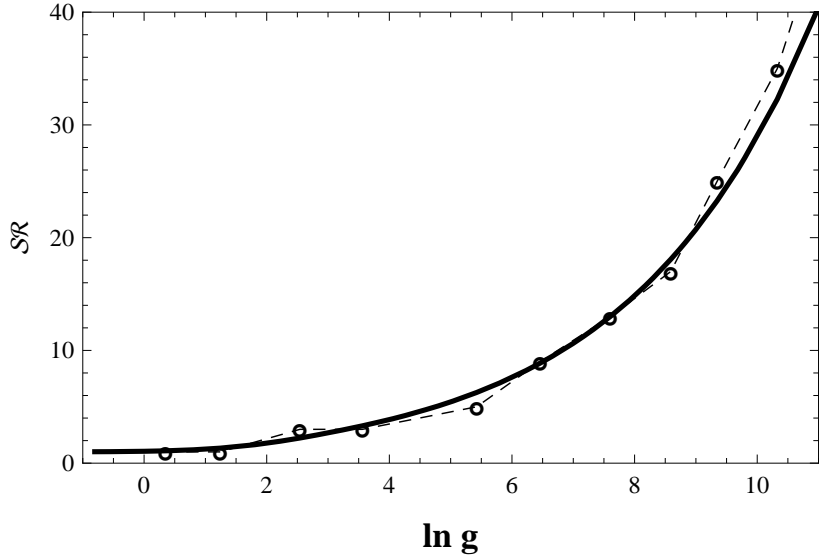}
\end{center}
 \caption{\label{fig2ffo65e3f:beh}The  effective Slater rank $SR=\textbf{R}/2$
  as a function of
   $\ln g$. The circles mark the results obtained by SR$= 2N_{l}-1$.
   }
 \end{figure}

To start with our analysis, we need    the solutions of
(\ref{radial}). As was shown by Taut  \cite{tautB}
   for a countably infinite set of   $g$ values  the closed-form wavefuntions
    can be derived.  $g=\sqrt{2}$ is the
smallest value at which an  exact ground-state wavefunction is known.
 Below this value no analytical solutions to the lowest eigenvalue of  (\ref{radial})
   are known  and  we have determined them numerically by the Rayleigh-Ritz
method. As was already discussed,  a qualitative insight into the
character of the partial-wave expansion  of $\psi$ can be provided
by examining the values of the collective occupancies. In Fig.\
\ref{fig25ee3f:beh} the numerically determined behaviour of
$\eta_{l}$   is presented
 for      $l=0-5$, where
 to  highlight the changes with $g$ a logarithmic
 coordinate has been used.
We may notice that   at $g\lesssim 1$  the  collective occupancy
with $l=0$
  is about  $1$. In that situation
 the   angular correlation effects may be neglected since one can expect
  that the terms higher than $l = 0$ in (\ref{partwave}) contribute very little.
  When  $g$ increases, more and more     collective occupancies with
different $l$  become  substantial,
  which  reflects the fact
that the larger is the value of $g$, the larger is the number of
partial waves  actively involved in the sum in (\ref{partwave}).
Interestingly enough, all the collective occupancies \emph{except}
for the largest one ($\eta_{0}$) exhibit a local maximum. The larger
is $l$, the larger is the value of $g$ at which the maximum occurs.
For large enough values of $g$ the situation changes essentially,
namely the clustering of collective occupancies is visible.
 It may have a general interest
to note at this point that at  $g \lesssim 7.4 $ ($\ln 7.4\approx
2$)
 the collective occupancies with $l> 1$ are vanishingly small,  whereas
   a Wigner molecule is expected to occur when $g$  exceeds a  value  $g_{cr}\simeq 2.8$ (See Fig.\ 5 in Ref.\ \cite{Puente},
    remember that $g=R_{w}\sqrt{2}$).
    Thus, in the incipient Wigner molecule regime,
     the angular correlations  come  mainly from the first angular level
     ($l=1$).
Further, in the range of $g$ in which  $||A_{l}||^2=\eta_{l}$ is
substantial, we  analyse  the dependence of the effective
 Schmidt number for    $A_{l}$, Eq.~(\ref{schaaa}), on
 $g$.
  Our numerical  results  are presented  in  Fig.\ \ref{fig2ffo65e3f1:beh} in terms of the  inverse  participation
ratio, $tr \bar{\rho_{l}}^2$$=\Omega_{l}$,  for   $l=0-3$ as a
function of $\ln g$. The results  obtained indicate nothing but  that
those  partial waves that contribute considerably to $\psi$
 have their radial parts $A_{l}$ being almost  uncorrelated functions.
   That is, $A_{l}(r_{1},r_{2})\sim\chi_{0,l}(r_{1})\chi_{0,l}(r_{2})$, as we have verified
   by solving   Eq.~(\ref{aa}) through a discretization technique. In other words,
   we have the result that  only the Schmidt orbitals with $(s=0,l)$ in
   (\ref{sc1}) are important and, in consequence, the total wavefunction   approaches  the form
\begin{eqnarray}\Psi(\zeta_{1},\zeta_{2})
\approx{\frac{1}{\sqrt{2}}}(\alpha(1)\beta(2)-\alpha(2)\beta(1))\times
\nonumber\\\sum_{l=0}^{N_{l}-1}\kappa_{0,l}
[\nu_{0,l}(\textbf{r}_{1})
\nu_{0,l}(\textbf{r}_{2})+\upsilon_{0,l}(\textbf{r}_{1})\upsilon_{0,l}(\textbf{r}_{2})]\label{sc},\end{eqnarray}
 where $N_{l}$ is the number of substantial partial waves.
The inspection of  (\ref{sc}) yields that it constitutes a sum of
$2N_{l}-1$ Slater determinants (here we recall  that
$\upsilon_{0,0}=0$). In the limit of $g\rightarrow\infty$,
  the values of  $\{\Omega_{l}\}_{l=0}$   saturate
to the same value $\Omega^{(\infty)}\simeq0.9634$, which reflects
the fact that at $g\rightarrow \infty$ the occupancies are
asymptotically independent of the angular momentum i.e.,
$\lambda_{s,0}^{g\rightarrow\infty}=\lambda_{s,1}^{g\rightarrow\infty}=...=\lambda_{s}^{(\infty)}$
($\rho_{0}^{g\rightarrow\infty}=\rho_{1}^{g\rightarrow\infty}=...
=\rho^{(\infty)}$).  We find this consistent with the properties of
the 3D Hooke's atom that were discussed in \cite{ocup1111}. What
follows further from the above is that in the limit
$g\rightarrow\infty$, the
 radial components  $\{A_{l}\}_{l=0}$ can differ from each other  only by their
 signs.
  We have verified that  $A_{l}^{g\rightarrow
\infty}=(-1)^{l}A^{(\infty)}$. Substitution of this  into
Eq.~(\ref{partwave}) gives
\begin{eqnarray}
\psi^{g\rightarrow \infty}(r,R)={{A}^{(\infty)}(r_{1},r_{2})\over
\sqrt{r_{1}r_{2}}}[{1\over 2 \pi} +{1\over
\pi}\sum_{l=1}^{\infty}{(-1)^{l}cos(l
\theta})].\label{3eff}\end{eqnarray}
Because ${(-1)^{l}cos(l \theta)}={cos[l (\theta-\pi)]}$,  the
suitable part of Eq.~(\ref{3eff}) can be recognized as a Dirac delta
distribution centred at $\pi$ i.e. $\delta(\theta-\pi)$\footnote {A
Fourier series expansion of $\delta(x-s)$, gives
$\delta(x-s)={1\over 2\pi}+{1\over
\pi}\sum_{l=1}^{\infty}cos[x(l-s)]$\cite{four}}. The
      angular correlations carry thus the electrons strictly at opposite  sides of the centre of the trap
     as $g\rightarrow\infty$, which corresponds to a perfect linear Wigner molecule.
In this limit the potential energy dominates over the electron
kinetic energy and the electrons localize to the classical positions
$r_{1,2}^{cl}={{1\over 2}({g\over 2}) ^{1\over 3}}$, $\theta=\pi$,
but always there exist   quantum fluctuations with respect to their
radial coordinates. The electrons oscillate about the equilibrium
geometry in accordance with a  harmonic approximation model
\cite{tautB,MatulisPeeters,Puente}. At $g\rightarrow\infty$ there is
no fluctuation in $\theta$, but the angular correlations are
extremely strong due to an arbitrary orientation of the molecular
axis, which coincides with the fact that the lowest-energy classical
configuration  is infinitely degenerate with respect to rotations
around the center of the trap.
 When it comes to
       pure radial correlations  appearing in the asymptotic regime  they are expected to be very poor,
        due to the closeness of the value of $\Omega^{(\infty)}$ to unity
        ($\Omega^{(\infty)}\simeq0.9634$). More precisely, we have found numerically that the ratio   of
${\lambda_{0}^{(\infty)}/ \lambda_{s}^{(\infty)}}$ is about
$~54,~2883,~154804$ for $s=1,2,3$, respectively, and it tends to
rapidly increase with further increasing $s$. In other words,  the
radial component appearing in Eq.~(\ref{3eff}) can indeed be
considered as almost a product function, which in turn implies  low
pure radial correlation effects. We stress that the limiting values
of the considered quantities have been obtained with the use of the
wavefunction  in the harmonic approximation, which is valid as
${g\rightarrow\infty}$.
 Before going further we want to  stress that in  none of  the well known papers  concerning
      the system under consideration  has  Eq.~(\ref{3eff}) appeared.
To gain further understanding  we next quantify   the degree of
entanglement by calculating the effective Slater rank
$SR=\textbf{R}/2$.
 An approximate value for $\textbf{R}$ can be obtained
with the help of  Eq.~(\ref{line}) by successively increasing the number of terms
in the sum appearing in it
  until the result   becomes stable to a desired  accuracy.
   In Fig.~\ref{fig2ffo65e3f:beh} the behaviour  of
    $SR$  determined by the procedure described above is shown.
We have checked  that     the results for the   linear entropy
 calculated with  Eq.~(\ref{line}) as  $\textbf{
L}=1-1/\textbf{R}$ completely agree with the results presented   in
Ref.\ \cite{kos1}, where they were calculated in a different way. The
circles in Fig.~\ref{fig2ffo65e3f:beh} represent the values of the
Slater rank obtained  from our earlier finding SR$=2N_{l}-1$ with
   $N_{l}$ being determined as was discussed at the end of  subsection \ref{pr}.
As one can see,  the behaviour of SR  coincides fairy well with that of  $SR$, which  justifies the approximation (\ref{sc}).

\section{Summary}\label{summ}

In conclusion, we have investigated the nature of the degeneracy in the
spectrum of the spatial RDM of the two-particle
wavefunction that is a function of distances $r$ and $R$  only. 
We have developed the results of  Ref.\ \cite{kos,sun} by  providing
the
 Schmidt  decomposition in the single real particle basis in parabolic
 coordinates.
In addition, we carried out a comprehensive study  of the singlet
ground-state of two
 interacting electrons
confined in an isotropic harmonic potential. As a general trend we
found that as $g$ increases, the number of partial waves that are
needed  to capture most of the electrons' correlation increases as
well.
  The Slater rank grows monotonically with
increasing $g$ and the bulk of the entanglement is mainly manifested
in the angular variables. At $g\rightarrow \infty$
  the spatial wavefunction factorizes asymptotically into a product of    radial and  angular
components, of which the former is   individually  little
correlated, but the latter is maximally correlated. Our calculations
have shown that only at $g\lesssim 1$ can the total wavefunction  be
well approximated by a single Slater determinant. In other words, in
this regime  the state under consideration can be regarded as weakly
entangled.

 It would be interesting to apply the tools presented in this Letter
to explore the entanglement properties of  two-particle systems
 with other interactions between the particles. This will be the topic of our further studies.

\bibliography{aipsamp}

\end{document}